\documentclass{PoS}

\usepackage{amsmath}
\usepackage{amssymb}
\usepackage{latexsym}

\newcommand{\ice}[1]{\relax}

\newcommand{\ed}{\end{document}}

\newcommand{\beq}{\begin{equation}}
\newcommand{\eeq}{\end{equation}}
\newcommand{\bea}{\begin{eqnarray}}
\newcommand{\eea}{\end{eqnarray}}

\newcommand{\ba}{\begin{array}}
\newcommand{\ea}{\end{array}}

\newcommand{\g}{\gamma}

\newcommand{\bc}{\begin{center}}
\newcommand{\ec}{\end{center}}

\newcommand{\re}[1]{(\ref{#1})}

\catcode`\@=11

\def\slash{\mathpalette\make@slash}
\def\make@slash#1#2{\setbox\z@\hbox{$#1#2$}%
  \hbox to 0pt{\hss$#1/$\hss\kern-\wd0}\box0}
\catcode`\@=12 % @ signs are no longer letters

\def\bbuildrel#1_#2^#3%
{\mathrel{\mathop{\kern 0pt#1}\limits_{#2}^{#3}}}

\newcommand{\sbz}{  }
\newcommand{\nnb}{\nonumber}
\newcommand{\as}{a_s}

\newcommand{\MSbar}{\ensuremath{\overline{\text{MS}}}}

\title{
Towards QCD running in 5 loops: quark mass anomalous dimension}

\ShortTitle{Towards QCD running in 5 loops: quark mass anomalous dimension}

\author{P.~A.~Baikov,\\
       Skobeltsyn Institute of Nuclear Physics, Lomonosov Moscow State University, 
      1(2), \\Leninskie gory, Moscow  119991, Russian Federation
      \\
        E-mail: \email{baikov@theory.sinp.msu.ru}}

\author{\speaker{K.~G.~  Chetyrkin} and J.~H.~K\"uhn,%
         \ice{\thanks{A footnote may follow.}}
\\
        Institut f\"ur Theoretische Teilchenphysik, Karlsruhe
  Institute of Technology (KIT), D-76128 Karlsruhe, Germany\\
E-mails : \email{konstantin.chetyrkin@kit.edu} \\ \hspace{1.4cm}\email{Johann.kuehn@kit.edu} }

\abstract{

We report  first results of an ongoing project devoted to the
analytical calculation of the QCD $\beta$-function and the quark mass 
anomalous dimension at the five loop level.
}

\FullConference{11th International Symposium on Radiative Corrections 
(Applications of Quantum Field Theory to Phenomenology) \\
		 22-27 September 2013\\
		 Lumley Castle Hotel, Durham, UK}

\begin{document}

\section{Introduction}

The method of the renormalization group (RG) \cite{Stueckelberg53,GellMann:1954fq,Bogolyubov:1956gh} 
is  of vital importance in modern quantum field 
theory. It is  enough  to  recall that the famous idea of the 
asymptotic freedom is based  on the RG concept of the running coupling 
constant.  The RG functions --- $\beta$-functions and various 
anomalous dimensions --- serve as  coefficients in the RG equations 
and  are expressed in terms of Feynman Integrals (FI's).  The 
complexity of these integrals grows drastically with the number of 
loops.

During last three decades or so there has been a tremendous progress
in our ability to compute analytically the RG functions. The progress
has been under way in, essentially, three directions.

i. General developments of our ability to deal with multiloop
FI's. These have been thoroughly documented by Vladimir Smirnov in his
bestseller books ``Evaluating Feynman integrals" and ``Feynman
Integral Calculus'' \cite{Smirnov:2004ym2,Smirnov:2006ry2} (see also
\cite{Smirnov:2002pj,Smirnov:2012gma2}). As a result two types of most
relevant for RG calculations FI's, namely, massless propagators  and
(completely) massive tadpoles (p- and m-integrals correspondingly)
can be calculated  (completely analytically) at the four loop level.

ii.  Invention of special  tools for significant simplifications of RG calculations. 
These include Infrared Rearrangement 
\cite{Vladimirov:1979zm,Chetyrkin:1980pr,Tarasov:2013zv} and  $R^*$-operation \cite{Chetyrkin:1984xa}.

iii. Continuous development of Computer Algebra Systems, with FORM \cite{Vermaseren:2000nd}
as  most prominent and indispensable tool.

The current state  of  art of (analytical) RG calculations can  be
summarized as follows: generic four-loop RG calculations are now    possible  
(see,  e.g.
\cite{Gorishnii:1991kd,gssql4,vanRitbergen:1997va,Schroder:2002re,Czakon:2006ss,Velizhanin:2011es,Bern:2013uka})
and  five-loop ones are  gradually getting ``feasible'' \cite{Baikov:2012zm,Eden:2012fe}.

In this talk we give the results of the first {\bf complete} calculations of
{\bf some of } QCD RG functions at five loops. These are the quark mass  anomalous
dimension   and the  anomalous dimension of the ghost field.
The latter is one  ingredient (among three)  necessary for
the construction of the five-loop contribution to the QCD $\beta$-function.

The precise evaluation of the quark mass anomalous dimension has
important implications. The Higgs boson decay rate into charm and
bottom quarks is proportional to the square of respective quark mass
at the scale of $m_H$ and the uncertainty from the presently unknown
5-loop terms in the running of the quark mass  is of order
$10^{-3}$.  This is comparable to  the precision advocated
for experiments e.g. at TLEP \cite{Gomez-Ceballos:2013zzn}. Similarly, the  issue of
Yukawa unification is affected  by precise  predictions for  the anomalous quark mass dimension.

\section{Preliminaries}
Our starting point is the QCD Lagrangian with $n_f$ quark  flavours written in terms of 
renormalized fields, coupling constant $g$ and quark massess $m_f$:
\begin{eqnarray}
{\cal L}_0 & = &
- \frac{1}{4} Z_3\, ( \partial _{\mu}A_{\nu} -  \partial _{\nu}A_{\mu})^2
- \frac{1}{2}g\, Z_1^{3g} \, ( \partial _{\mu}A^a_{\nu} -  \partial _{\nu}A^a_{\mu})
\,  ( A_{\mu} \times A_{\nu})^a 
\nonumber
\\
&-&
 \frac{1}{4} g^2\, Z^{4g}_1\, ( A_\mu \times A_\nu)^2
-  \frac{1}{2  \xi_L }  ( \partial _\nu A_\mu)^2
+ Z^c_3\, \partial_\nu \bar c  \, (\partial_\nu c )
+ g\, Z_1^{ccg} \, \partial^\mu \bar c \, (A \times   c  )
\label{lagr:2}
\\
&+&
Z_2\sum_{f=1}^{n_f} 
\bar \psi^f  \mathrm{i}  \slash{ \partial } \, \psi^f
+
g Z^{\psi\psi g}_1
\sum_{f=1}^{n_f} 
\bar \psi^f   \slash{A}\, \psi^f
- Z_{\psi\psi}  \sum_{f=1}^{n_f} 
 m_f \,\bar \psi^f  \, \psi^f
\ice{
Z_2\sum_{f=1}^{n_f} 
\bar \psi^f ( \mathrm{i}  \slash{ \partial } 
 + g Z^{\psi\psi g}_1 Z_2^{-1}\slash{A} - Z_{\psi\psi} Z_2^{-1} m_f )\, \psi^f
}
\nonumber
{},
\end{eqnarray}
with bare  gluon, quark and ghost fields related to the renormalized ones as follows:
\begin{equation} 
A^{a \mu}_0 = \sqrt{Z_3}\ A^{a \mu},
\ \
\psi^f_0  = \sqrt{Z_2}\ \psi^f_0,
\ \
c^{a}_0   = \sqrt{Z_3^{c}}\ c^{a} 
{}.
 \end{equation} 
The   vertex Renormalization Constants (RCs)
 \begin{equation} 
Z^V_1, \ \ \ V\in \{\mathrm{3g,\ 4g,\  c  c  g  , \ \psi  \psi g}\}
{}
 \end{equation} 
are to  be chosen to renormalize 3-gluon, 4-gluon, ghost-ghost-gluon, 
quark-quark-gluon vertex functions respectively. 
The  Slavnov-Taylor identities allows one to express  all   vertex
RCs in terms of wave function RCs and  an  independent  charge 
RC,  $Z_g = \frac{ \displaystyle  g_0 }{ \displaystyle  g}$:
 \begin{eqnarray} 
Z_\xi &=& Z_3, 
\label{WI:xi}
\\
Z_g &=& \sqrt{Z_1^{4g}} \,  (Z_3)^{-1}, \ \ 
\label{WI:4g}
\\
Z_g &=& Z_1^{3g} (Z_3)^{-3/2}, \ \ 
\label{WI:3g}
\\
Z_g &=& Z_1^{ccg} (Z_3)^{-1/2} (Z_3^c)^{-1}, \ \ 
\label{WI:ccg}
\\
Z_g &=& Z_1^{\psi\psi g} (Z_3)^{-1/2} (Z_2)^{-1}
\label{WI:qqg}
{}.
 \end{eqnarray} 

Within the commonly accepted $ \overline{\mbox{MS}} $ scheme RCs are  independent of 
dimensional parameters (masses and momenta) and can be represented
as follows
 \beq
%Z(h) = 1 + \sum_{n=1}^\infty \frac{z^{(n)}(h)}{ \epsilon ^n}
Z(h) = 1 + \sum_{i,j}^{1 \le j \le i} Z_{ij}  \frac{h^i}{ \epsilon^j}
\label{}
{},
 \eeq
where $h = g^2/(16 \pi^2)$ and the parameter $ \epsilon $ is related to the continuous   
space time dimension $D$  via $D= 4 - 2 \epsilon $.
Given a RC $Z(h)$,  the  corresponding anomalous dimension is defined as
 \begin{equation} 
\gamma(h) = -\mu^2\frac{\mathrm{d} \log Z(h)}{\mathrm{d} \mu^2}
%=h \frac{ \partial  z^{(1)}(h)}{ \partial  h}
= \sum_{n=1}^\infty  Z_{n,1}\, n\, h^n
= -\sum_{n=0}^\infty (\gamma)_n \, h^{n+1} 
\label{anom:dim:generic}
{}.
\end{equation}

The anomalous dimension of the charge $h$ is conventionally  referred to as ``QCD $\beta$-function''; equations
(\ref{WI:4g}-\ref{WI:qqg})  imply that all four expressions   in the Table below

\begin{table}[h]
\bc
\begin{tabular}{ |c|c|c|c| }
\hline
\multicolumn{4}{ |c| }{$\beta(h) = $} \\
\hline
$2\gamma_1^{ccg} - 2 \  \gamma_3^{c} -  \, \gamma_3 $ &
$ 2\gamma_1^{\psi\psi g} - 2 \  \gamma_2 -  \, \gamma_3$ &
$ 2 \, \gamma_1^{3g} - 3 \, \gamma_3 $ &
$ \gamma_1^{4g} - 2 \, \gamma_3 $\
\\ \hline
\end{tabular}
\ec
\caption{Four different representation   the QCD $\beta$-function.} 
\end{table}
\noindent 
can be  used to find the QCD $\beta$-function $\beta(h)$. In real calculations  only the first or  the second
possibility is usually employed. 

To calculate the  quark mass anomalous dimension,  $\gamma_m$, one needs to 
find the so-called quark mass renormalization constant, 
$Z_{m}$, which is defined as the ratio of the bare and renormalized
quark masses, viz.  
\beq
Z_m = \frac{m_f^0}{m_f} = \frac{Z_{\psi\psi}}{Z_2}
\label{Zm}
{}.
\eeq
The final formula for $\g_m$ reads
\beq
\g_m =  \g_{\psi\psi}  - \g_2
{}.
\eeq

\section{Five-loop running of the ghost field}

As a first step  towards five-loop QCD  $\beta$-function we have computed  the anomalous dimension of the ghost  field
\beq
\gamma^{c}_3 = -\sum_{i=0}^{\infty} (\gamma^{c}_3)_i 
h^{i+1}
{}.
\eeq
The anomalous dimension is known  up to four loops from the  works \cite{Chetyrkin:2004mf,Czakon:2004bu}. 
The new five-loop coefficient reads
(in the Feynman gauge):
\begin{eqnarray}
(\gamma^{c}_3)_{4}&=& 
%\nonumber\\
% 
-\frac{193301287}{2048} 
-\frac{19562145}{128}  \sbz \zeta_{3}
-\frac{2060829}{128}  \,\zeta_3^2
+\frac{1101573}{16}  \sbz \zeta_{4}
\nonumber\\
&{+}&
\frac{66632427}{128}  \sbz \zeta_{5}
-\frac{36327825}{256}  \,\zeta_{6}
-\frac{140900823}{512}  \,\zeta_{7}
%zero == 0
\nonumber\\
&{+}& \, n_f 
\left[
\frac{633704171}{27648} 
+\frac{5166473}{144}  \sbz \zeta_{3}
+\frac{233519}{64}  \,\zeta_3^2
-\frac{764949}{32}  \sbz \zeta_{4}
\right.
\nonumber\\
&-&
\left.
\frac{32902291}{384}  \sbz \zeta_{5}
+\frac{4123825}{128}  \,\zeta_{6}
+\frac{14425075}{384}  \,\zeta_{7}
%zero == 0
\right]
\nonumber\\
&{+}& \, n_f^2
\left[
-\frac{1326547}{3456} 
-\frac{1739167}{864}  \sbz \zeta_{3}
-\frac{2659}{6}  \,\zeta_3^2
+\frac{13485}{8}  \sbz \zeta_{4}
+\frac{8074}{9}  \sbz \zeta_{5}
-\frac{16775}{12}  \,\zeta_{6}
%zero == 0
\right]
\nonumber\\
&{+}& \, n_f^3
\left[
-\frac{342895}{7776} 
-\frac{1211}{18}  \sbz \zeta_{3}
-\frac{5}{2}  \sbz \zeta_{4}
+\frac{284}{3}  \sbz \zeta_{5}
%zero == 0
\right]
\nonumber
+ \, n_f^4
\left[
\frac{65}{108} 
+\frac{20}{27}  \sbz \zeta_{3}
-\frac{4}{3}  \sbz \zeta_{4}
%zero == 0
\right]
\label{gcc5l}
\end{eqnarray}

Numerically ({\large $\as \equiv \frac{\alpha_s}{\pi} \equiv 4\, h$} ):
\[
\gamma^{c}_3 (n_f=3)  = \frac{3}{8}\left( \as + 2.4375\, \as^2  + 4.8867\, \as^3  
+ 19.980 \,\as^4  \,  {+ \, 122.246 \,\as^5 }\,\right)
{}.
\]
For generic  $n_f$:
\bea
\gamma^{c}_3   &=& \frac{3}{8}\left\{
\as
+  \as^2\left(
3.063 - 0.208 n_f
\right)
+  \as^3\left(
10.556 - 1.768 n_f - 0.0405 n_f^2
\right)
\right.
\nnb
\\
&{}& 
\left.
\hspace{9mm}
+ \ \as^4\left(
49.325 - 10.957 n_f + 0.36562 n_f^2  + 0.0087 n_f^3
\right)
\right.
\nnb
\\
&{}&
\left.
{{
 +\as^5\left(
283.632 - 70.979 n_f + 5.498 n_f^2  + 0.0769 n_f^3
-0.000128038 n_f^4
\right)}
}
\right\}
\nnb
{}.
\eea
It is instructive to observe that significant cancellations between $n_f^0$ and $n_f^1$ terms  for
the values of $n_f$ around 3 or so persist also at five-loop order.

%\section{five-loop running of the  ghost-ghost-gluon vertex }

\section{Five-loop quark mass anomalous dimension}

The quark mass anomalous dimension is known   to four loops from the  works \cite{Chetyrkin:1997dh,Vermaseren:1997fq}.
Our result for the five-loop coefficient  in
\beq
\gamma_{m} = -\sum_{i=0}^{\infty} \ (\gamma_{m})_i \ h^{i+1}
\eeq
reads (note that $\gamma^{m}$ is  a gauge independent quantity):
\begin{eqnarray}
(\gamma_m)_{4} &=& 
\nonumber
%\frac{pp1}{4^5}
\frac{99512327}{162} 
 + \frac{46402466}{243}  \sbz \zeta_{3}
 + 96800  \,\zeta_3^2
 - \frac{698126}{9}  \sbz \zeta_{4}
\nonumber\\
&{}& \hspace{1.2cm}
 - \frac{231757160}{243}  \sbz \zeta_{5}
 + 242000  \,\zeta_{6}
 + 412720  \,\zeta_{7}
%zero == 0
\nonumber\\
&{+}& \, n_f 
\left[
-\frac{150736283}{1458} 
 - \frac{12538016}{81}  \sbz \zeta_{3}
 - \frac{75680}{9}  \,\zeta_3^2
 + \frac{2038742}{27}  \sbz \zeta_{4}
\right.
\nonumber
\\
&{}& 
\left. 
\hspace{1.2cm}
 + \frac{49876180}{243}  \sbz \zeta_{5}
 - \frac{638000}{9}  \,\zeta_{6}
 - \frac{1820000}{27}  \,\zeta_{7}
%zero == 0
\right]
\nonumber\\
&{+}& \, n_f^2
\left[
\frac{1320742}{729} 
 + \frac{2010824}{243}  \sbz \zeta_{3}
 + \frac{46400}{27}  \,\zeta_3^2
 - \frac{166300}{27}  \sbz \zeta_{4}
 - \frac{264040}{81}  \sbz \zeta_{5}
 + \frac{92000}{27}  \,\zeta_{6}
%zero == 0
\right]
\nonumber\\
&{+}& \,
\fbox{$
\normalsize
 n_f^3
\left[
\frac{91865}{1458} 
 + \frac{12848}{81}  \sbz \zeta_{3}
 + \frac{448}{9}  \sbz \zeta_{4}
 - \frac{5120}{27}  \sbz \zeta_{5}
%zero == 0
\right]
+ \, n_f^4
\left[
-\frac{260}{243} 
 - \frac{320}{243}  \sbz \zeta_{3}
 + \frac{64}{27}  \sbz \zeta_{4}
\right]
%\Biggr\}
$}
{}.
\nonumber
\end{eqnarray}
Note that the boxed terms are in full agreement with predictions made on 
the basis of the $1/n_f$ method 
in \cite{Ciuchini:1999wy,Ciuchini:1999cv}.

In  numerical form $\g_m$ reads
\bea
\nonumber
\g_m =  &-& a_s - a_s^2 \left(4.20833 - 0.138889 n_f\right)
\\ \nonumber
&-&
a_s^3  \left(19.5156 - 2.28412 n_f - 0.0270062 n_f^2 \right)  
\\ \nonumber 
&-&
a_s^4 \left(98.9434 - 19.1075 n_f + 0.276163 n_f^2  + 0.00579322 n_f^3 \right)
\\
&-&
a_s^5 \left(
559.71 - 143.6\, n_f + 7.4824\, n_f^2  + 0.1083\, n_f^3  - 0.00008535\, n_f^4
\right)
{}
\label{N[gm5qcd]}
\eea
\ice{
\bea
\nonumber
\g_m =  &-& a_s - a_s^2  (4.20833 - 0.138889 n_f)
\\ \nonumber
&-&
a_s^3  (19.5156 - 2.28412 n_f - 0.0270062 n_f^2 )  
\\ \nonumber 
&-&
a_s^4  (98.9434 - 19.1075 n_f + 0.276163 n_f^2  + 0.00579322 n_f^3 )
\\
&-&
559.71 + 143.6\, n_f - 7.4824\, n_f^2  - 0.1083\, n_f^3  + 0.00008535\, n_f^4
{}
\label{N[gm5qcd]}
\eea
}
and
 \bea
\nonumber                               
\g_m \bbuildrel{=\!=\!=}_{n_f = 3}^{}
&-& \as - 3.79167 \,\as^2  - 12.4202 \,\as^3  - 44.2629 \,\as^4  - 198.907 \,\as^5
 ,\\ \nonumber                                            
g_m \bbuildrel{=\!=\!=}_{n_f = 4}^{}
&-& \as - 3.65278 \,\as^2  - 9.94704 \,\as^3  - 27.3029 \,\as^4  - 111.59 \,\as^5
,\\  \nonumber 
\ g_m \bbuildrel{=\!=\!=}_{n_f = 5}^{}                                                      
&-& \as - 3.51389 \,\as^2  - 7.41986 \,\as^3  - 11.0343 \,\as^4  - 41.8205 \,\as^5
 ,\\      
\g_m \bbuildrel{=\!=\!=}_{n_f = 6}^{}
&-&  \as - 3.375   \,\as^2  - 4.83867 \,\as^3  + 4.50817 \,\as^4  + 9.76016 \,\as^5
\label{gm5:nf:3-6}
{}.
\eea
Inspection of eqs. \re{gm5:nf:3-6} shows quite moderate growth of the
series in $\as$ appearing in the quark mass anomalous dimension at
various values of active quark flavours (recall that even for scales
as small as 2 GeV $\as \equiv \frac{\alpha_s}{\pi} \approx 0.1$).

\ice{
In[6]:= Coll[gm5/.a->as/4/.Zrule/.qcd//N,{as,nf}]

                   2
Out[6]= -1. as + as  (-4.20833 + 0.138889 nf) + 
 
       3                                      2
>    as  (-19.5156 + 2.28412 nf + 0.0270062 nf ) + 
 
       4                                     2                3
>    as  (-98.9434 + 19.1075 nf - 0.276163 nf  - 0.00579322 nf ) + 
 
       5                                    2              3
>    as  (-559.707 + 143.686 nf - 7.48238 nf  - 0.108318 nf  + 
 
                       4
>       0.0000853589 nf )

}

\section{Technical tools}

As is well-known  evaluation of any $L$-loop anomalous
dimension in the \MSbar-scheme  can be reduced, with the help of the $R^*$-operation \cite{Chetyrkin:1984xa}, to the
evaluation of some $L-1$-loop massless propagators \cite{Chetyrkin:1996ez}. 
In our case $L=5$ and we need to be able effectively  compute a host of four-loop massless  propagators (that is  p-integrals).
These, in turn, can be  reduced to 28 master integrals. The reduction
is based on evaluating  
sufficiently many terms of the $1/D$ expansion \cite{Baikov:2005nv} of
the corresponding coefficient functions \cite{Baikov:1996rk}. The master integrals are known analytically from
\cite{Baikov:2010hf,Lee:2011jt}.

Note that  all our calculations   have  been performed
on a SGI ALTIX 24-node IB-interconnected cluster of 8-cores Xeon
computers using  parallel  MPI-based \cite{Tentyukov:2004hz} as well as thread-based 
\cite{Tentyukov:2007mu} versions  of FORM
\cite{Vermaseren:2000nd}.

\section{Conclusions}

Unfortunately, at the moment it is not possible to take
self-consistently into account our five-loop result for $\g_m$ for the
quark mass running: this  requires  the  knowledge of the five-loop QCD $\beta$-function.
The latter problem is under calculation in our group.

K.G.C. thanks  J. Gracey and  members of the DESY-Zeuthen theory  seminar 
for  usefull discussions.  

This work was supported by
the Deutsche Forschungsgemeinschaft in the
Sonderforschungsbereich/Transregio
SFB/TR-9 ``Computational Particle Physics'',  
by  RFBR (grant 11-02-01196).

\providecommand{\href}[2]{#2}\begingroup\raggedright\endgroup

\end{document}